\documentclass[prb,aps,twocolumn,draft,tightenlines,%
showpacs,floatfix]{revtex4}
\usepackage{psfig}
\usepackage{amssymb}
\usepackage{amsmath}

\begin{document}

\title{Comment on "Spin-split two-dimensional electron gas perturbed
  by intense terahertz laser field"}

\author{J. L. Cheng}
\affiliation{Hefei National Laboratory for Physical Sciences at
  Microscale, University of Science and Technology of China, Hefei,
  Anhui, 230026, China}
\affiliation{Department of Physics, University of
  Science and Technology of China, Hefei, Anhui, 230026, China}
\altaffiliation{Mailing Address.}


\date{\today}
\begin{abstract}
  We show that the time-dependent wavefunctions which serve as
basis of the whole paper of Xu [Phys. Rev. B {\bf 70}, 193301 (2004)]
are incorrect and point out the right ones.

\end{abstract}

\maketitle
\newcommand{\bfk}[1]{\mathbf {#1}}

In a recent paper\cite{xu} Xu addressed
the electron spins of the two-dimensional electron gas
in the presence of an intense terahertz laser field.
All the quantities
such as the Green function, the density of 
states and the number of the particle he calculated are based on the 
time dependent wavefunction
({\em i.e.} Eq. (2) in Ref. 1) 
of the electron Hamiltonian $H(t)$ (Eq. (1) in Ref. 1). 
However that paper deserves a comment as the wavefunction 
given by Xu is {\em incorrect}.

It is easy to check that the
wavefunction given by Xu\cite{xu} {\em does not} satisfy the
Schr\"odinger equation
\begin{widetext}
\begin{eqnarray}
(i\hbar\partial_t - H(t))\Psi_{n\bfk{k}\sigma}(\bfk{R}, t) &=&
-\frac{\hbar\omega r_0k_{\alpha}\sin(\omega t)}{\sqrt{2}}e^{i\bfk{k}\bfk{r}}\psi_n(z)
\binom{\sigma - i\sigma \frac{k_y-ik_x}{k}}{i+\frac{k_y-ik_x}{k}}e^{-i[E_{n\sigma}(k)
+E_{em}]t/\hbar}e^{i\gamma
  \sin(2\omega t)}e^{ir_0(k_x+\sigma k_{\alpha})[1-\cos(\omega
  t)]}\nonumber\\
&\neq& 0\ .
\end{eqnarray}
\end{widetext}
Therefore his remaining results are questionable.

The right wavefunctions expanded in the Floquet
space was given in our recent paper.\cite{cheng} Differing from the wavefunctions
given by Xu where two spin branches $\Psi_{n{\bf k}+}({\bf R},t)$ and
$\Psi_{n{\bf k}-}({\bf R},t^\prime)$ are always orthogonal to each other
for any $t$ and $t^\prime$, for the right time-dependent wavefunctions
these two branches are only orthogonal when $t=t^\prime$.  
Therefore for  the right wavefunctions,
one gets not only the diagonal terms of the Green functions, but also
the off-diagonal terms which indicate the correlations between
the two spin branches. These correlations result in new spin polarizations
which are absent in Xu's paper.
More results in detail can be found in our
paper.\cite{cheng}

\end{document}